\documentclass[conference]{IEEEtran}
\IEEEoverridecommandlockouts

\usepackage{cite}
\usepackage{amsmath,amssymb,amsfonts}
\usepackage{algorithmic}
\usepackage{graphicx}
\usepackage{textcomp}
\usepackage{xcolor}
\usepackage{comment}
\usepackage{makecell}  
\usepackage{tabularx} 
\usepackage{xspace}

\usepackage{paralist} 
\usepackage{caption}
\usepackage{multirow}
\usepackage{booktabs,threeparttable}
\usepackage{color}
\usepackage{threeparttable}
\usepackage{colortbl}
\usepackage[autostyle=true,english=american]{csquotes}
\usepackage{subcaption}
\usepackage{url}
\usepackage{balance}
\usepackage{fancyhdr}
\usepackage{multirow}
\usepackage{hyperref}
\usepackage{framed}
\usepackage[most]{tcolorbox}
\usepackage{tikz}
\usepackage{tikz}
\def\BibTeX{{\rm B\kern-.05em{\sc i\kern-.025em b}\kern-.08em
    T\kern-.1667em\lower.7ex\hbox{E}\kern-.125emX}}

\definecolor{hellgrau}{RGB}{245,245,245}

\definecolor{verygood}{RGB}{44,123,182}
\definecolor{good}{RGB}{171,217,233}
\definecolor{ok}{RGB}{255,255,191}
\definecolor{bad}{RGB}{253,174,97}
\definecolor{verybad}{RGB}{215,25,28}

\newcommand\rel{\textit{Relate}\xspace}
\newcommand\sub{\textit{Subtask}\xspace}
\newcommand\dup{\textit{Duplicate}\xspace}
\newcommand\clo{\textit{Clone}\xspace}
\newcommand\dep{\textit{Depend}\xspace}
\newcommand\epi{\textit{Epic}\xspace}
\newcommand\blo{\textit{Block}\xspace}
\newcommand\inc{\textit{Incorporate}\xspace}
\newcommand\cau{\textit{Cause}\xspace}
\newcommand\rep{\textit{Replace}\xspace}
\newcommand\spl{\textit{Split}\xspace}

\def\colorModel{RGB} 
\def\cca#1{%
    \pgfmathsetmacro\compA{200+(#1*306)-(475*#1*#1)}      
    \pgfmathsetmacro\compB{24.6+(#1*787)-(691*#1*#1)} 
    \pgfmathsetmacro\compC{15.7+(#1*511)-(334*#1*#1)}      
  \edef\clrmacro{\noexpand\cellcolor[\colorModel]{\compA,\compB,\compC}}\clrmacro
  {#1}%
  }
  
\newtcolorbox[auto counter]{findingbox}[3][]{
arc=0.5mm,
lower separated=false,
fonttitle=\bfseries,
colbacktitle=gray!20,
coltitle=gray!50!black,
enhanced,
attach boxed title to top left={xshift=6cm,
        yshift=-2mm},
colframe=gray!50!black,
colback=gray!20,
overlay={
\node[draw=gray!50!black,thin,
fill= gray!10,rounded corners=0.5mm, 
yshift=0pt, 
xshift=-0.1cm, 
left, 
text=gray!50!black,
anchor=east,
font=\bfseries] 
at (frame.north east) {#3};},
overlay={

at (frame.north east) {#3};},
title=#2 \thetcbcounter,#1}

\begin{document}
\title{Automated Detection of Typed Links \\ in Issue Trackers}
\author{
    \IEEEauthorblockN{ 
        Clara Marie L{\"u}ders, 
        Tim Pietz, and
        Walid Maalej
    }
    \IEEEauthorblockA{
        \textit{Universit\"at Hamburg},
        Hamburg, Germany \\
        E-Mail: clara.marie.lueders@uni-hamburg.de, tim.pietz@uni-hamburg.de, walid.maalej@uni-hamburg.de
    }
}

\maketitle


\begin{abstract}

Stakeholders in software projects use issue trackers like JIRA to capture and manage issues, including requirements and bugs. 
To ease issue navigation and structure project knowledge, stakeholders manually connect issues via links of certain types that reflect different dependencies, such as Epic-, Block\mbox{-}, Duplicate-, or Relate- links.
Based on a large dataset of~15 JIRA repositories, we study how well state-of-the-art machine learning models can automatically detect common link types.
We found that a pure BERT model trained on titles and descriptions of linked issues significantly outperforms other optimized deep learning models, achieving an encouraging average macro F1-score of~0.64 for detecting 9 popular link types across all repositories (weighted F1-score of~0.73). 
For the specific Subtask- and Epic- links, the model achieved top F1-scores of~0.89 and~0.97, respectively. 
Our model does not simply learn the textual similarity of the issues. In general, shorter issue text seems to improve the prediction accuracy with a strong negative correlation of~-0.70. 
We found that Relate-links often get confused with the other links, which suggests that they are likely used as default links in unclear cases. 
We also observed significant differences across the repositories, depending on how they are used and by whom. 


\end{abstract}


\IEEEpeerreviewmaketitle

\section{Introduction}
Software development teams use issue trackers to manage and maintain software products and their requirements.
Popular issue trackers in practice are Bugzilla\footnote{\url{https://www.bugzilla.org}}, GitHub Issues\footnote{\url{https://github.com}}, and JIRA\footnote{\url{https://www.atlassian.com/software/jira}}, with common features like issue creation and tracking, communication around issues, release planning, issue triaging, and issue dependency management.
As a central entity, issues have many properties including their type~\cite{Montgomery:2022} (e.g. requirement, feature request, bug report, or task), priority (e.g. low, high, critical), and assignee.
Issue trackers and issue management, in general, have been the focus of software engineering and requirements engineering research for over a decade, with promising automation approaches, e.g., on issue type prediction (and correction)~\cite{Perez:ICPC:2021, Herzig:ICSE:2013, Winkler:RE:2019, Hey:2020:RE, Merten:2016:RE}, priority (and escalation) prediction~\cite{Malhotra:CONF:2021, Fang:IEEE:2021, Montgomery:2017:RE, Fitzgerald:2011:RE}, or triaging and assignment~\cite{Stanik:2018:ICSME, Jeong:FSE:2009}.

Issues are often interconnected via \textit{links}~\cite{Li:APSEC:2018, Montgomery:2022}, which enables stakeholders to capture the relationships between the issues and structure the overall project knowledge.  
Depending on the issue tracker and the project, these links can have different types.
Popular link types are \rel for capturing a general relation; \sub and \epi for capturing issue hierarchies; as well as \dep or \blo links for capturing causal or workflow dependencies.
Also, \dup links are particularly popular in open source projects, where many stakeholders and users independently report issues that might be a duplication. This specific link type has attracted much attention from Software Engineering research in recent years~\cite{Deshmukh:ICSME:2017,He:ICPC:2020,Wang:ICSE:08,Lazar:MSR:2014}.

Issue linking is an important part of issue management and software engineering in general. 
Research has found that linking helps reduce issue resolution time~\cite{Li:APSEC:2018} and prevent software defects~\cite{Rempel:IEEE:2017}.
Missing or incorrect links can particularly be problematic for requirements analysis and release planning~\cite{Thompson:MSR:2016}.
For instance, missing \dep or \blo links to issues assigned to a specific release might be crucial for that release. 
Similarly, missing duplicate links might lead to missing additional context  information~\cite{Cavalcanti:CSMR:2010}, which can particularly be relevant for testing. 

As issue trackers are getting more central for documenting almost all project activities, particularly in agile projects, and with the evolution of software products over time, a project can easily get thousands of issues. 
Each new issue might thus have hundreds of thousands of potentially relevant links. 
Correctly identifying and connecting issues quickly become difficult, time-consuming, and error-prone~\cite{Fucci:ESEM:2018, Lucassen:REW:2017}.
This is even worse in popular issue trackers like those of Apache, RedHat, or Qt, which are open to users and other stakeholders and which may contain hundreds of thousands of issues.
The problem becomes also more complex when link types are taken into account: the issue creator or maintainer does not only have to decide if a link between two issues exists but also what the correct link type is.




To tackle this problem, we report on a novel study, which aims at automatically predicting the issue links including their specific type. Our work is enabled by the advances in deep learning technology (particularly for natural language texts) as well as by the availability of large issue datasets. 
Our work has two main contributions.
First, we compare multiple state-of-the-art, end-to-end deep-learning models to predict and classify the issue links. 
Our evaluation shows that a BERT model using the description and title of the two involved issues is by far the most accurate to classify typed links.
There are barely any limitations to the applicability of the model, as the issue types are not predefined but learned from the data. 
The results are very encouraging, particularly since up to~13 link types are predicted.  
Second, we compare the properties of the various repositories and link types and we analyze potential correlations with the performance of the model.
The results reveal insights on link types confusion as well as on how to design and apply link prediction tools in practice. 
In addition to the recently published dataset of~15 JIRA repositories \cite{Montgomery:2022}, we share our code and analysis scripts to ease replication.\footnote{\url{https://github.com/RegenKordel/LYNX-TypedLinkDetection}}
    
The remainder of the paper is structured as follows. 
Section~\ref{sec:research-method} outlines our research questions, method, data, and the various classification models used. 
Section~\ref{sec:BERT-results} presents the performance results of the BERT model for typed link prediction across~15 different issue repositories.
In Section~\ref{sec:Repo-LT-analysis}, we compare the repositories and the link types with the factors that influence the prediction performance.
In Section~\ref{sec:discussion}, we discuss our findings, their limitations, possible link prediction optimization strategies, and implications for different stakeholders.
Finally, we discuss related work in Section~\ref{sec:rel-work} and conclude the paper in Section~\ref{sec:conclusion}.

\section{Research Setting}\label{sec:research-method}

We outline our research questions and data as well as the research method and machine learning models investigated.

\subsection{Research Questions and Data}

Motivated by a) the importance of the typed link prediction in issue management~\cite{Li:APSEC:2018, Rempel:IEEE:2017, Thompson:MSR:2016}, b) recent work on issue duplicate prediction~\cite{He:ICPC:2020},  c) recent advances of transformer-based machine learning technique for natural language processing BERT~\cite{Devlin;arxiv;2018}, as well as d) the availability of large issue  datasets~\cite{Montgomery:2022}, we studied typed links prediction in issue trackers focusing on two main questions: 

\begin{itemize}
    \item \textbf{RQ1.} How well can state-of-the-art machine learning predict the different link types in issue trackers?
    \item \textbf{RQ2.} What are possible explanations for the performance differences between repositories and link types and what can we learn from the differences?
\end{itemize}

For answering the research questions, we used a large dataset originally consisting of 16 public JIRA repositories~\cite{Montgomery:2022}.
This dataset perfectly matches our research goals. 
Not only is JIRA one of the most popular tools for issue management in practice\footnote{\url{https://enlyft.com/tech/products/atlassian-jira}}\footnote{\url{https://www.datanyze.com/market-share/project-management--217/jira-market-share}}\footnote{\url{https://www.slintel.com/tech/bug-and-issue-tracking}}; it is also well-known for its support for various link types, which can be customized depending on the project needs. 
Other available datasets, e.g.~of Bugzilla~\cite{Lazar:MSR:2014}, only focus on the specific link type \dup and have intensively been used in software engineering literature (particularly Mining Software Repositories) so far~\cite{He:ICPC:2020, Deshmukh:ICSME:2017, Xiao:ISSRE:2020}.
Bugzilla also supports multiple link types, such as \rel and \dep, but JIRA repositories usually include a larger variety of link types.

Table~\ref{tab:DescStats} summarizes the analyzed issue repositories.
The table shows the year of creation, number of issues, number of links, unique link types, coverage, number of projects, and the share of cross-project links. 
Coverage represents the number of issues having at least one link.
The share of cross-project links is the share of links that connect issues of different projects in a repository. 
We bolded the minimum and maximum for each metric across all repositories in the table.

The investigated repositories are heterogeneous in the reported properties.
The number of reported issues ranges from~1,867 to~1,014,926, while the number of links ranges from~44 links in Mindville to~255,767 links in Apache.
The repositories also vary in terms of link types: Mindville uses~4 unique link types while Jira (corresponding to the development of the JIRA issue tracker) and Apache use~16 unique link types.
As~44 training points are too low for a stratified split, we excluded Mindville from the analysis.
On average, the coverage is about~36\% meaning that~36\% of all issues are part of a link. 
The coverage ranges from~4.0\% in Mindville to~54.9\% and~53.7\% in Hyperledger and Mojang respectively.
Except in Jira, RedHat, and MongoDB links rarely cross project boundaries.
The majority (95\% on average) of links are between issues of the same project.
The Jira repository shows a high coverage among its projects~(46.7\%).

\begin{table}
\setlength\tabcolsep{2.5pt}  
\centering
\captionsetup{skip=2pt}
\caption{Studied JIRA repositories in alphabetical order.}\label{tab:DescStats}
\begin{tabularx}{\columnwidth}{X rrrrrrr}
\multicolumn{8}{c}{\footnotesize{\makecell{Columns: Documented Link Types (\#Types);\\ percentage of issues with a link (\%Cov.);\\ the percentage of links for issues from two different subprojects. (\%CP)}}}\\
\hline
\textbf{Repo.}                 & \textbf{Year} & \textbf{\#Issues} &  \textbf{\#Links} & \textbf{\#Types}  &  \textbf{\#Pro.} &  \textbf{\%Cov.}   & \textbf{\%CP}\\
\hline 
\rowcolor{hellgrau}Apache      & \textbf{2000} &  \textbf{1014926} & \textbf{255767}   &  \textbf{16}      & \textbf{646}               &  28.5\%           &          5.2\% \\
Hyperledger                   & \textbf{2016}  &   28146           &  16304            &          8        & 32                & \textbf{ 54.9\%}           &          4.6\% \\
\rowcolor{hellgrau}IntelDAOS   & \textbf{2016} &    9474           &   2599            &         11        & \textbf{2}                 &  30.8\%           &          3.8\% \\
JFrog                        & 2006   &   15535           &   3229            &         10        & 10                &  28.6\%           &          8.2\% \\
\rowcolor{hellgrau}Jira      & 2002   &  274545           &  99819            &         \textbf{16}        & 30                &  46.7\%           &         \textbf{43.2\%} \\
JiraEcosystem               & 2004    &   41866           &  11398            &         14        & 101               &  33.0\%           &          6.8\% \\
\rowcolor{hellgrau}MariaDB  & 2009    &   31229           &  14618            &          8        & 11                &  44.5\%           &          2.5\% \\
Mindville                   & 2015    &   2134  &     \textbf{44}   &  \textbf{4}       & 7                 &  \textbf{4.0\%}   &          4.6\% \\
\rowcolor{hellgrau}Mojang   & 2012    &   420819           &  215527            &         5         & 8                 &  53.7\%           &          5.4\% \\
MongoDB                    & 2009     &   137172           &  63821            &         14        & 27                &  45.2\%           &         19.1\% \\
\rowcolor{hellgrau}Qt      & 2005     &  148579           &  40105            &          11        & 21                &  30.2\%           &          6.9\% \\
RedHat                  & 2001        &  353000           & 119669            &         15        & 241               &  39.2\%           &         23.5\% \\
\rowcolor{hellgrau}Sakai  & 2004      &   50550           &  19803            &          8        & 53                &  42.4\%           &          \textbf{1.4\%} \\
SecondLife                & 2007      &    \textbf{1867}           &    631            &          6        & \textbf{2}                 &  39.9\%           &          2.4\% \\
\rowcolor{hellgrau}Sonatype   & 2008  &   87284           &   4465            &         11        & 5                 &  7.0\%            &          1.5\% \\
Spring                      & 2003    &   69156           &  14462            &          7        & 80               &  25.6\%           &         10.0\% \\
\hline
\textbf{Total} & - &  2686282  &  882261  & 164 &  1276 & - & -   \\
\textbf{Median} & - &  59853  &  15461  & 10.5 & 24 & 36.1\% & 5.3\%  \\
\hline
\end{tabularx}
\end{table}


When analyzing the data, we observed that a link might point to a private issue.
As we have no further information about private issues, we excluded these links from the analysis. 
We also removed multi-links, i.e. when two public issues were part of multiple links with different types.
This affected~1.1\% of all public links.
We also simplify the analysis by ignoring link direction.

\begin{table*}[ht]
\setlength\tabcolsep{3.75pt}  
\centering
\captionsetup{skip=2pt}
    \caption{Popular link types and their usage shares in percent across the studied JIRA repositories.}
    \label{tab:RepoLtFreq}
    \begin{tabularx}{0.93\textwidth}{X rrrrrrrrrrr | r}
    \hline
\textbf{Repository} &  Relate & Duplicate & Subtask &   Clone &   Block &  Depend &    Epic &  Split & Incorporate & Bonfire Testing &  Cause & \textbf{Coverage} \\
\hline
\rowcolor{hellgrau}Apache             &   28.3 &      10.1 &    32.8 &   1.7 &   6.1 &    5.1 &   4.9 &   0.0 &         4.1 &             0.0 &   1.2 &     94.3 \\
Hyperledger        &   17.2 &       3.9 &    27.6 &   2.9 &   8.2 &      / &  39.6 &   0.5 &           / &             0.01 &     / &    100.0 \\
\rowcolor{hellgrau}IntelDAOS          &   39.3 &       9.7 &    10.5 &   1.5 &  25.5 &      / &     / &     / &           / &               / &     / &     86.6 \\
JFrog              &   27.4 &      19.9 &    36.0 &   0.8 &     / &    7.9 &     / &     / &         1.4 &               / &     / &     93.5 \\
\rowcolor{hellgrau}Jira               &   63.8 &      21.7 &     2.5 &   2.9 &   1.0 &    0.2 &     / &   0.2 &         2.5 &             0.2 &   1.8 &     96.6 \\
JiraEcosystem      &   22.9 &      15.3 &    20.0 &   1.8 &   5.9 &    1.1 &  24.2 &   1.2 &         1.8 &             0.9 &   3.9 &     99.1 \\
\rowcolor{hellgrau}MariaDB            &   51.1 &       9.4 &     6.1 &     / &  13.0 &      / &   6.4 &   0.2 &         7.9 &               / &   6.0 &    100.0 \\
Mindville          &   43.2 &      38.6 &       / &  15.9 &   2.3 &      / &     / &     / &           / &               / &     / &    100.0 \\
\rowcolor{hellgrau}Mojang             &    9.5 &      90.0 &       / &   0.3 &   0.1 &      / &     / &     / &           / &             0.1 &     / &    100.0 \\
MongoDB            &   39.9 &      13.5 &     1.4 &   0.3 &     / &   22.9 &  15.9 &   1.2 &           / &               / &   1.7 &     96.7 \\
\rowcolor{hellgrau}Qt                 &   22.4 &      10.6 &    24.4 &   0.1 &   0.03 &   15.6 &  13.5 &   6.7 &           / &               / &     / &     93.4 \\
RedHat             &   25.9 &       4.9 &    20.8 &  15.4 &  15.2 &      / &     / &   0.1 &         8.9 &               / &   2.6 &     94.0 \\
\rowcolor{hellgrau}Sakai              &   49.0 &       9.3 &    17.0 &   4.8 &   0.03 &   13.0 &     / &     / &         6.7 &             0.01 &     / &    100.0 \\
SecondLife         &   29.5 &         / &    49.8 &   7.6 &     / &    4.4 &     / &     / &         2.2 &               / &     / &     93.5 \\
\rowcolor{hellgrau}Sonatype           &   40.0 &       7.7 &    30.1 &     / &     / &    3.6 &   0.2 &   0.1 &           / &             8.1 &   5.3 &     95.0 \\
Spring             &   47.7 &      12.1 &    13.4 &   0.1 &     / &   12.1 &  11.3 &     / &           / &               / &     / &     96.7 \\
\hline
Mean               &   \textbf{34.8} &      \textbf{18.4} &    \textbf{20.9} &   4.0 &   7.0 &    8.6 &  14.5 &   \textbf{1.1} &         4.4 &             \textbf{1.3} &   3.2 &     96.2 \\
Standard Deviation &   \textbf{14.3} &      \textbf{21.5} &    \textbf{13.8} &   5.4 &   8.1 &    7.2 &  \textbf{12.5} &   2.1 &         3.0 &             3.0 &   1.9 &      3.7 \\
\hline
\end{tabularx}
\end{table*}


Out of the 31 unique link types found in the dataset, most appear in less than half of all repositories. 
Uncommon link types only represent a small share of the links.
Table~\ref{tab:RepoLtFreq} shows the frequencies of link types that are used by at least 7 repositories.
The highest variance can be observed for \rel, \dup, \sub, and \epi. 
We focus our study on link types that have a share greater than~1\% in the respective repository. 
For instance, Qt's \clo makes only~0.1\% in all of Qt's links and thus will not appear in the results table in Section~\ref{sec:BERT-results}.
Additionally, to ensure a minimum amount of comparability and generalizability across the repositories, we exclude link types from the analysis if their total share across all repositories is less than~2\% (which leads to excluding \spl and \textit{Bonfire Testing} from the analysis).
As a result, our analysis focuses on the following common link types: \rel, \dup, \sub, \dep, \epi, \clo, \inc, \cau, and \blo. 
A detailed description of the link types can be found in~\cite{Montgomery:2022}.
We particularly note the difference between \dup and \clo authors: \dup links represent accidentally created reports of the same issue whereas \clo links are automatically created when a user uses the ``clone'' feature of JIRA.
The link types \blo and \dep are similar.
But as some issue trackers, such as Apache and JiraEcosystems use both types in parallel, we refrained from merging them.


\subsection{Research Method and Machine Learning Models}

For answering RQ1, we built, trained, and compared multiple deep learning end-to-end models that predict the link type or the absence of a link for an issue pair. 
This includes a BERT model, which recently attracted much attention in the NLP community.  
For RQ2, we focused on the top-performing model and looked for factors that correlate with the prediction performance, in particular, properties of the repositories studied, the link types, as well as properties of the linked issues.


We focused our research on a minimal model input, consisting of the title and description. 
These are rather universal issue properties (independently of the tracker and project at hand) and are usually available at the issue creation time.
This means that we did not  study the usage of additional features for the prediction, such as the issue type or status which might influence the prediction performance. Such features might also  contain information about the label (link types) which could create a bias in the model (e.g., the resolution for duplicate issues could be  ``duplicate'').

Our labels consist of the link types as used in the repository by the stakeholders.
We also added \textit{non-links} as a label  by randomly selecting closed issue pairs that do not have the resolution property ``duplicate''. 
For instance, the model trained on Qt had 8 labels \rel, \sub, \dep, \epi, \dup, \spl, \rep, and \textit{non-link}.
We kept the number of  \textit{non-links} in the experiment always equal to the mean counts of the other labels: to get enough data points so that the model can learn but also not too many data points which could  create an imbalance. 
We then randomly split the data for each repository into a training~(80\%) and a test set~(20\%).
For the model validation, we took~20\% of the training data, resulting in 64/16/20 train-validate-test sets.
We also stratified the data by link type.
We used the test set only for the evaluation.


For selecting the model's architecture, we checked the Software Engineering and the Natural Language Processing (NLP) literature. 
In the NLP community, BERT~\cite{Devlin;arxiv;2018} and DistilBERT~\cite{Sanh:arxiv:2019} have recently attracted much attention for various NLP tasks.
In Software engineering, different approaches for duplicate detection have been recently suggested, with two main model architectures.  
Single-Channel architectures~\cite{Deshmukh:ICSME:2017} take the word representation (word2vec, fasttext, or GLOVE) and encode each issue separately with a siamese network, consisting of a CNN, LSTM, and a feed-forward neural network.
Then, the two encoded issues are fed into another forward neural network to determine if one issue is a duplicate of the other.
Dual-Channel architectures\cite{He:ICPC:2020} use the word representations of two issues (which are two matrices of word embeddings) to construct a tensor by putting the two matrices on top of each other. 
Thus, in this architecture, the two issues are encoded jointly.

We experimented with BERT, DistilBERT, the Single-Channel, and Dual-Channel architectures using FastText and Word2Vec.
DistilBERT is a smaller transformer model trained to imitate BERT through knowledge distillation.
It retains most of BERT's general language understanding capabilities while using~40\% fewer model weights.
We included DistilBERT in our comparison because we thought a model with fewer parameters might perform better on smaller repositories.

In preliminary experiments, we found that BERT outperformed all other models in all setups:  
BERT outperformed DistilBERT by an average of~0.05 F1-score,
the Single-Channel models by an average of~0.21, and the Dual-Channel models by an average of~0.26.
The code and results of all  evaluated models are included in the replication package. 
In the remainder of the paper, we focus on discussing the results of  BERT as best performing model.



We concatenated the title and description of both issues into one string and used this as input for the BERT model.
Then, we tokenized this input string with the tokenizer of bert-base-uncased / distilbert-base-uncased, which is a trained WordPiece tokenizer.
We truncated the token sequence at 192 tokens with a \textit{longest first} strategy. That is, if possible, we kept all tokens and otherwise truncate the longer of the two issues first.
The \texttt{[CLS]} token output of the BERT model was then fed into a dense layer classification head,  which predicts the label of the link.
For the training, we chose AdamW and use the default learning rate of~$5e^{-5}$ and weight decay of~0.1.
We ran the training on NVIDIA Tesla K80 GPUs using the largest possible batch size that fit into the GPU memory, which was~96 with BERT and~128 with DistilBERT.
We trained for~30 epochs and evaluated the model on the validation set after every epoch, reporting only on the model with the lowest validation F1-score.

We report the F1-score per common link type and the macro and weighted averages of the model per repository.
Detailed tables for recall and precision are included in the replication package, without specific results which are worth discussing in the paper.
We also present the normalized confusion matrix for each repository.
We chose the conservative macro F1-score as our primary metric to evaluate performance.
Weighted averages tend to overestimate model performance, as models tend to predict instances of majority classes better since there are more data points to learn from.
Furthermore, neither using class weights nor SMOTE as strategies to counter the class imbalance showed any improvements.
As we are unaware of any baselines of link type detection in JIRA, we created a baseline by using TF-IDF with Random Forest and SVM and compared this against our results.
We balanced both baseline classifiers with class-weights and report the macro F1-score of a 5-fold cross-validation.

\begin{table*}[ht!]
\setlength\tabcolsep{2.5pt}  
\centering
\caption{F1-scores for predicting issues links and their types across the studied JIRA repositories. A bluer color indicates that the F1-score is closer to 1 while a reddish color indicates a result closer to 0. }\label{tab:LTinRepoStats}
\captionsetup{skip=2pt}
\begin{tabularx}{\textwidth}{X|rrrrrrrrrr|rr|rr}
\multicolumn{13}{c}{\footnotesize{\makecell{RF: Random Forest Baseline (Macro F1), SVM: Support Vector Machine Baseline (Macro F1)}}}\\
\hline       
\textbf{Repository} &  Relate & Duplicate & Subtask & Depend & Clone & Incorporate &  Epic & Block & Cause &  Non-Link & \textbf{Macro F1}& \textbf{Weight. F1} & \textbf{RF} & \textbf{SVM} \\
\hline
Apache        &      \cca{0.64} &        \cca{0.49} &  \cca{ 0.91} &       \cca{0.46} &      \cca{0.61} &      \cca{0.55} &  \cca{0.97} &      \cca{0.52} &      \cca{0.34} &      \cca{0.76} &        \cca{0.56} &       \cca{0.70} & \cca{0.12} & \cca{0.13}  \\
Hyperledger   &       \cca{0.70} &       \cca{0.37} &  \cca{ 0.89} &           / &   \cca{ 0.80} &          / &  \cca{ 0.97} &    \cca{0.62} &           / &  \cca{ 0.85} &      \cca{0.74} &  \cca{ 0.85} & \cca{0.30} & \cca{0.27}  \\
IntelDAOS     &      \cca{0.69} &        \cca{0.44} &  \cca{ 0.82} &           / &  \cca{ 0.86} &          / &            / &     \cca{0.7} &           / &        \cca{0.48} &      \cca{0.72} &      \cca{0.68} & \cca{0.37} & \cca{0.35}  \\
JFrog         &        \cca{0.57} &        \cca{0.51} &  \cca{ 0.95} &       \cca{0.45} &            / &  \cca{0.0} &            / &          / &           / &      \cca{0.72} &        \cca{0.48} &      \cca{0.66} & \cca{0.25} & \cca{0.27}  \\
Jira          &  \cca{ 0.88} &      \cca{0.71} &  \cca{ 0.93} &           / &      \cca{0.61} &    \cca{0.64} &            / &          / &      \cca{0.37} &  \cca{ 0.86} &      \cca{0.73} &  \cca{ 0.82}& \cca{0.27} & \cca{0.23}  \\
JiraEcosystem &      \cca{0.62} &        \cca{0.57} &  \cca{ 0.89} &  \cca{0.19} &        \cca{0.59} &     \cca{0.29} &  \cca{ 0.98} &      \cca{0.41} &      \cca{0.38} &       \cca{0.60} &        \cca{0.53} &      \cca{0.71} & \cca{0.17} & \cca{0.17}  \\
MariaDB       &      \cca{0.76} &       \cca{0.37} &  \cca{ 0.86} &           / &            / &    \cca{0.68} &  \cca{ 0.97} &    \cca{0.66} &       \cca{0.52} &      \cca{0.76} &       \cca{0.7} &      \cca{0.72} & \cca{0.27} & \cca{0.33}  \\
Mojang        &       \cca{0.70} &  \cca{ 0.97} &            / &           / &            / &          / &            / &          / &           / &  \cca{ 0.96} &  \cca{ 0.88} &  \cca{ 0.95} & \cca{0.48} & \cca{0.48}  \\
MongoDB       &      \cca{0.73} &        \cca{0.46} &  \cca{ 0.85} &     \cca{0.69} &            / &          / &  \cca{ 0.97} &          / &      \cca{0.36} &      \cca{0.72} &      \cca{0.72} &      \cca{0.72} & \cca{0.31} & \cca{0.25}  \\
Qt            &        \cca{0.56} &        \cca{0.43} &  \cca{ 0.93} &     \cca{0.75} &            / &          / &  \cca{ 0.96} &          / &           / &  \cca{ 0.81} &      \cca{0.66} &      \cca{0.71} & \cca{0.27} & \cca{0.29}  \\
RedHat        &      \cca{0.63} &       \cca{0.29} &   \cca{ 0.90} &           / &  \cca{ 0.83} &    \cca{0.77} &            / &    \cca{0.63} &        \cca{0.40} &      \cca{0.72} &      \cca{0.63} &       \cca{0.70} & \cca{0.25} & \cca{0.21}  \\
Sakai         &      \cca{0.72} &       \cca{0.39} &  \cca{ 0.88} &       \cca{0.48} &      \cca{0.65} &      \cca{0.57} &            / &          / &           / &      \cca{0.78} &      \cca{0.64} &      \cca{0.68} & \cca{0.20} & \cca{0.22}  \\
SecondLife    &      \cca{0.73} &            / &  \cca{ 0.87} &        \cca{0.40} &         \cca{0.40} &  \cca{0.0} &            / &          / &           / &  \cca{ 0.85} &        \cca{0.52} &      \cca{0.74} & \cca{0.37} & \cca{0.30}  \\
Sonatype      &      \cca{0.68} &       \cca{0.33} &  \cca{ 0.91} &       \cca{0.42} &            / &          / &            / &          / &      \cca{0.26} &      \cca{0.73} &        \cca{0.46} &      \cca{0.65} & \cca{0.21} & \cca{0.21}  \\
Spring        &      \cca{0.73} &       \cca{0.38} &  \cca{0.82} &       \cca{0.48} &            / &          / &  \cca{0.99} &          / &           / &      \cca{0.74} &      \cca{0.62} &      \cca{0.69} & \cca{0.24} & \cca{0.28}  \\
\hline
Mean          &      \cca{0.69} &        \cca{0.48} &  \cca{0.89} &       \cca{0.48} &      \cca{0.67} &      \cca{0.44} &  \cca{0.97} &      \cca{0.59} &      \cca{0.38} &      \cca{0.76} &      \cca{0.64} &      \cca{0.73} & \cca{0.27} & \cca{0.27}  \\
Standard Dev. &    0.08 &    0.17 &    0.04 &   0.15 &    0.14 &      0.28 &    0.01 &   0.10 &   0.07 &    0.11 &    0.11 &    0.08 & & \\
\hline
\end{tabularx}
\end{table*}

\section{Prediction Performance}\label{sec:BERT-results}
\subsection{Overall Prediction Results}


Table~\ref{tab:LTinRepoStats} shows the common link types and their respective F1-scores for each repository. 
We also present the mean and standard deviation per link type as well as the macro and weighted F1-score per repository.
Overall, the model macro F1-score achieves 0.64 on a multi-class problem with a median of 7 classes per repository. 
The weighted F1-score goes up to~0.73.   
The recall goes from 0.47 for Sonatype up to 0.89 for Mojang. The precision ranged from 0.46 for Sonatype up to 0.86 for Mojang.
The results are far better for all repositories than the Random Forest and SVM baseline with TF-IDF. Both baselines achieve an average F1-score of~0.27.

As expected, the results differ across repositories.
JFrog only has a macro F1-score of~0.48, while Mojang has a macro F1-score of~0.88.
Some of the variances across the repositories can be explained by the size of the training set. For instance, JFrog, SecondLife, and Sonatype each contain less than 5k links in total, whereas Mojang contains roughly 200k links.
We also observe that the performance of the model differs per link type.
The model detects \sub and \epi links consistently with a top performance, while the prediction of \dup, \dep, and \cau seem to be less accurate. 
The other link types \dep, \clo, \inc show mixed results. \textit{Non-Link} shows top performance for all but one repository.



\begin{figure}[]
\centering
\includegraphics[scale=0.35]{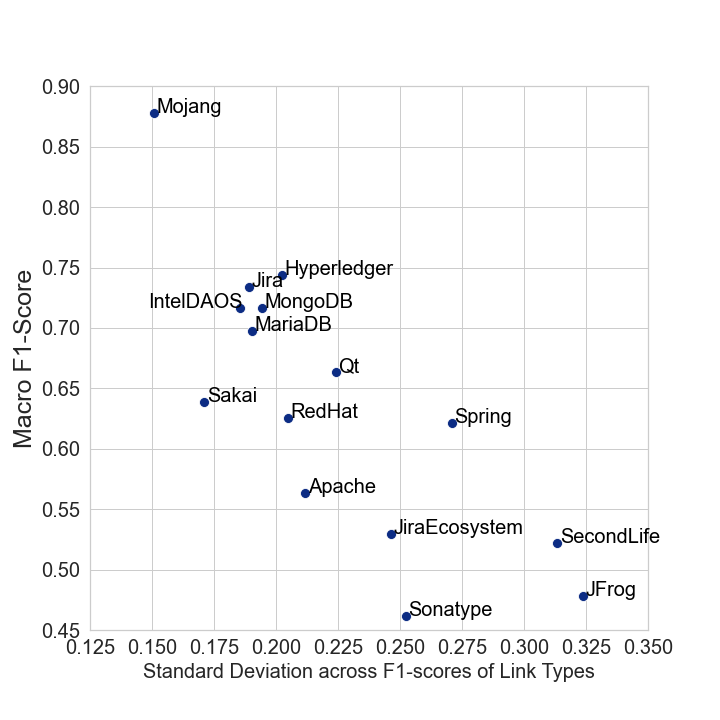}
\caption{Performance per repository according to macro F1-score and standard deviation for the studied link types.}
\label{fig:RepoPerformance}
\end{figure}

Figure~\ref{fig:RepoPerformance} plots the macro F1-score against the standard deviation of the F1-scores across the link types.
A higher standard deviation means that the model performs well for some but not for all link types. 
A low standard deviation means that the model performs similarly for all  types. 
Mojang, with a lot of available training data and only three different predicted link types (\dup, \rel, and \textit{non-link}) performs best (highest macro F1-score with lowest standard deviation).
The next cluster consists of Hyperledger, IntelDAOS, Jira, MariaDB, and MongoDB: with macro F1-scores ranging from~0.70 to~0.74 and a standard deviation around~0.19.
Then, Qt, RedHat, and Sakai perform slightly worse: their macro F1-scores lie between~0.63 and~0.66, while their standard deviation is less than~0.225.
The last cluster consists of Apache, JFrog, JiraEcosystems, Secondlife, Sonatype, and Spring. These repositories, all with lower coverage, either have a macro F1-score less than~0.60 or a higher standard deviation.
The case of Apache is particularly interesting as it has the highest number of links and issues, and one of the highest number of predicted classes.
It also contains the largest number of projects~(646), which indicates some internal heterogeneity.

\begin{tcolorbox}[colback=gray!5!white,colframe=gray!75!black, size=small]
\underline{\textbf{Finding 1.}}
A general BERT model applied to issue titles and descriptions predicts the typed links with a promising mean  macro F1-score of~0.64 across~15 repositories.
Some repositories show a top performance while the model achieved moderate performance for others.
The key difference seems to be the link coverage.
\end{tcolorbox}



\subsection{Individual Link Types and Confusion Analysis}

\begin{figure*}[]
\centering
\includegraphics[scale=0.135]{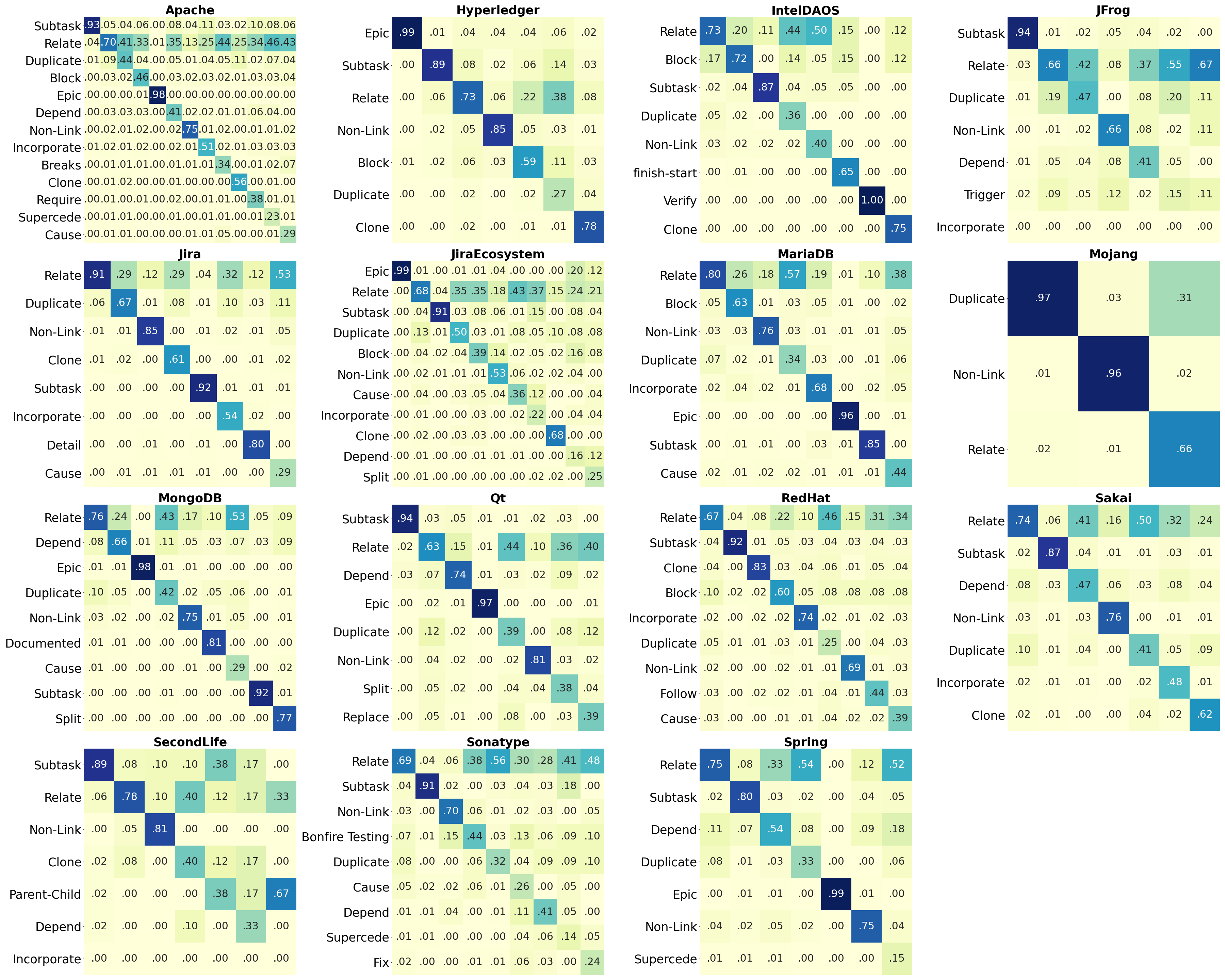}
    \caption{Normalized confusion matrices for each repository. Rows are sorted by the support of link type, meaning the majority class is always the first row. The columns are in the same order as the rows. 
    }
\label{fig:LTConfMat}
\end{figure*}

In the next step, we took a closer look at the various link types. 
Figure~\ref{fig:LTConfMat} shows the confusion matrices for each repository, ordered by the number of data points per link type in the test set.
We identified which link types were confusing to the model and thus consequently mislabeled.
Overall, \rel is often the majority class and commonly predicted by the model for other link types.
This sometimes happens for \dup but on a much smaller scale.
We also observe that \clo and \dup are well distinguished by the model.

\subsubsection{\rel Links}
These links are provided by default in a JIRA installation. They are thus used in all repositories. 
They are usually one of the largest classes. 
The model often mistakes other link types as \rel links. 
The F1-score ranges from~0.56 up to~0.88.
It is predicted fairly well with an average macro F1-score of~0.69.

\subsubsection{\dup \& \clo Links}
We originally assumed that the model will struggle to distinguish \clo  from \dup links and vice versa.
Surprisingly, the existence of the class \clo did not confuse the model.
They were rarely mistaken for each other, which  can be observed in the confusion matrices of Apache, Hyperledger, IndelDAOS, Jira, JiraEcosystem, RedHat, and Sakai.
This suggests that they exhibit a difference that the model can recognize.
\dup links have rather mixed F1-scores, ranging from~0.29 to~0.97. Only in Jira and Mojang \dup links were better classified with~0.71 and~0.97 F1-scores, respectively.
The other repositories achieved a macro F1-score up to~0.50.
In contrast to \dup issues that are usually created independently by different stakeholders, \clo links are usually intentionally created via the ``Clone Issue'' feature of JIRA. 
These links seem to be easier detectable by the model with F1-scores ranging from~0.40 to~0.86.

\subsubsection{\sub \& \epi Links}
The model showed top scores for classifying \sub and \epi links, with an average F1-score of~0.89 and~0.97 respectively.
\sub links ranging from~0.82 up to~0.95 and \epi links ranging from~0.97 up to~0.99, both also have a low standard deviation.
\sub and \epi links have an extra section or field in an issue view in JIRA and are not grouped under the ''Issue Links`` section as for the other link types.
Surprisingly, the model was able to differentiate \sub links from \epi links, although the performance for \sub was slightly lower if the repository includes the type \epi.

\subsubsection{\dep, \inc, \blo, \& \cau Links}
All of these link types show either  varying or rather low performances.
\dep links' performance ranged  from~0.19 to~0.75.
This could be explained by the link type share. Qt, with the best performance, has approx.~16\% of its links as \dep link types, followed by MongoDB, with approx.~23\% \dep links.
When the performance is low, either  \dep is not used much in the repository (i.e. the absolute count is less than~300) or it is confused with \rel (for Spring).

The performance for predicting \inc links also varied with a standard deviation of~0.28, and macro F1-scores ranging from~0.00 up to~0.77.
In JFrog and SecondLife, the number of training examples was less than~50 (44 and~14 respectively), explaining the~0.00 in both rows.
In cases where the performance is high, there were at least~1000 training examples in the repository.

\blo links had a moderate performance with a standard deviation of~0.10 and macro F1-scored from~0.41 up to~0.70.
Even though three repositories frequently use this link type, it only has an average of~4.8\% share across all repositories.
It seems that the model only struggles to distinguish this link type from \rel but not from other -- semantically rather similar -- link types, such as \dep or \cau.
We can also observe that \blo and \dep are used separately: either \blo is used or \dep -- except in Apache and JiraEcosystem.
Finally, we observe that the performance of \blo is high if the repository uses this type in more than~10\% of its links.

\cau links have a very low average share and low performance, ranging from~0.26 up to~0.52.
MariaDB had the best performance, but even there the model was unable to discern this link type from \rel.
The reason for this might simply be that a causal relation is harder to detect while simultaneously not having enough example data.

\subsubsection{Non-Links} We observed an encouraging to top  performance for predicting non-links, up to~0.96 in Mojang, except for IntelDAOS.  
Overall the model did not struggle to identify links from non-links.
The case of Mojang, which only has~3 link types to predict, is an indicator of how well the model can perform as a ``linked issues'' vs. ``non-linked issues'' detector.

\begin{tcolorbox}[colback=gray!5!white,colframe=gray!75!black, size=small]
\underline{\textbf{Finding 2.}}
The BERT model is overall accurate at distinguishing \textit{non-links} from links.
The structural link types \sub and \epi perform very well across all repositories. \dup does not perform well overall, but the model is able to distinguish \clo links from \dup links.
\dep, \inc, and \blo can be distinguished when they are frequently used.
\cau corresponds to the lowest prediction performance.\\
\rel seems to be the most confusing for the model as it is likely a jack-of-all-trades type used to label a link when a stakeholder is unsure which type fits.
\end{tcolorbox}

\section{Analysis of Differences}\label{sec:Repo-LT-analysis}

In this section, we explore possible  causes for the model performance differences across the repositories and link types.

\subsection{Repository Analysis}
We calculated the Pearson correlation and $p$-value of different repository properties listed in Table~\ref{tab:DescStats} to find out what properties correlate with the macro F1-scores of the model.
We also correlated the number of unique issue creators, reporters, assignees, and the total number of unique users.

\begin{table}[]
\centering
\captionsetup{skip=2pt}
\caption{Correlation coefficient and $p$-Value of macro F1-scores of the classifier to properties of the repositories.}\label{tab:RepoPropCorr}
\begin{tabular}{lrr}
\hline
\textbf{Properties} &     \textbf{Correlation} &      \textbf{$p$-Value} \\
\hline
\rowcolor{hellgrau}\#Issues         &  0.0937 &  0.7397\\
\#Links          &  0.3523 &  0.1978 \\
\rowcolor{hellgrau}\#Projects       & -0.2109 &  0.4505 \\
\rowcolor{hellgrau}\#PredictedTypes & -0.5039 &  0.0554 \\
\%Coverage       &  \textbf{0.7500}  &  \textbf{0.0013} \\
\rowcolor{hellgrau}\%CrossProject   &  0.2439 &  0.3811 \\
\#Total Users         &  0.4419 &  0.0991 \\
\rowcolor{hellgrau}\#Assignees       & -0.1567 &  0.5771 \\
\#Creators      &  0.4374 &  0.1030 \\
\rowcolor{hellgrau}\#Reporters & 0.4435 &  0.0978 \\
Assignee-Issue-Ratio       &  \textbf{0.5208}  &  \textbf{0.0465} \\

\rowcolor{hellgrau}\#Age & -0.4525 &  0.0903 \\
\hline
\end{tabular}
\end{table}

Table~\ref{tab:RepoPropCorr} shows the correlation of the properties with the macro F1-score.
The model's performance seems independent of the number of issues in the issue tracker. 
As expected, we observe a strong positive correlation with the coverage, an average positive correlation with the number of links, and a fairly strong negative correlation with the number of link types and the age of the issue tracker. 
A small negative correlation also exists for the number of projects in the repository.
Only the correlation with coverage is significant.

Moreover, we observe positive correlations with the number of total users, creators, and reporters, and a negative correlation with the number of assignees.
Although none of them are significant, it is interesting that the number of assignees correlates negatively with the prediction performance.
We thus calculated the number of issues per assignee in a repository, which turns out to have a significant, fairly strong, positive correlation with the prediction performance.
It is worth noting that, given the rather limited statistical power of the 15 studied repositories, it is not surprising that most correlations are not significant. 

\begin{tcolorbox}[colback=gray!5!white,colframe=gray!75!black, size=small]
\underline{\textbf{Finding 3.}}
Coverage and number of issues per assignee show strong and statistically significant correlations with the macro F1-score to predict typed links. 
A higher coverage can indicate that stakeholders place more value on linking. The more issues an assignee is responsible for, the more homogeneous a repository (and the linking) is likely to become.
\end{tcolorbox}

\subsection{Link Type Analysis}
We calculated the TF-IDF vectors of the title and description of the involved issues and used cosine similarity to calculate the similarity of the issue pairs on a word basis.
We also looked at the length of the textual descriptions of linked issues (word count of title and description of issue pairs) as well as the difference in length of two linked issues.
We report the results for the common types in the dataset.

\begin{table}[ht]
\setlength\tabcolsep{2.5pt}  
\centering
\captionsetup{skip=2pt}
    \caption{Median cosine similarity scores of the texts of linked issues on text-token-level across the JIRA repositories.}
    \label{tab:RepoLtCossim}
    \begin{tabularx}{\columnwidth}{X  rrrrrrrrrr}
    \hline
\textbf{Repo} &  Rel. &  Dup. &  Sub. &  Dep. &  Clo. &  Inc. &  Epic &  Blo. &  Cau. &  NL.\\
\hline
\rowcolor{hellgrau}Apache        &    0.14 &       0.29 &     0.00 &    0.08 &   0.90 &         0.07 &  0.00 &   0.03 &   0.09 &      0.00 \\
Hyperledger   &    0.33 &       0.44 &     0.21 &      / &   0.95 &           / &  0.06 &   0.31 &     / &      0.00 \\
\rowcolor{hellgrau}IntelDAOS     &    0.18 &       0.29 &     0.07 &      / &   0.99 &           / &    / &   0.11 &     / &      0.00 \\
JFrog         &    0.37 &       0.38 &     0.00 &    0.28 &   1.00 &         0.27 &    / &     / &     / &      0.02 \\
\rowcolor{hellgrau}Jira          &    0.89 &       0.44 &     0.06 &    0.49 &   0.92 &         0.40 &    / &   0.29 &   0.34 &      0.03 \\
JiraEcosystem &    0.25 &       0.48 &     0.00 &    0.15 &   0.94 &         0.13 &  0.00 &   0.13 &   0.24 &      0.00 \\
\rowcolor{hellgrau}MariaDB       &    0.25 &       0.32 &     0.06 &      / &     / &         0.13 &  0.00 &   0.22 &   0.13 &      0.04 \\
Mojang        &    0.24 &       0.19 &       / &      / &   0.37 &           / &    / &   0.19 &     / &      0.00 \\
\rowcolor{hellgrau}MongoDB       &    0.17 &       0.31 &     0.17 &    0.24 &   1.00 &           / &  0.00 &     / &   0.13 &      0.00 \\
Qt            &    0.24 &       0.30 &     0.21 &    0.12 &   0.91 &           / &  0.00 &   0.20 &     / &      0.00 \\
\rowcolor{hellgrau}RedHat        &    0.21 &       0.36 &     0.17 &      / &   1.00 &         0.00 &    / &   0.10 &   0.16 &      0.00 \\
Sakai         &    0.24 &       0.39 &     0.15 &    0.18 &   0.89 &         0.19 &   / &   0.12 &    / &      0.00 \\
\rowcolor{hellgrau}SecondLife    &    0.25 &         / &     0.17 &    0.15 &   0.58 &         0.00 &    / &     / &     / &      0.03 \\
Sonatype      &    0.27 &       0.40 &     0.00 &    0.14 &     / &           / &  0.34 &     / &   0.21 &      0.00 \\
\rowcolor{hellgrau}Spring        &    0.20 &       0.30 &     0.00 &    0.16 &   0.30 &           / &  0.00 &     / &     / &      0.00 \\
\hline
\textbf{Mean} & 0.28 & 0.35 & \textbf{0.09} & 0.20 & \textbf{0.83} & 0.15 & \textbf{0.05} & 0.17 & 0.19 & 0.01 \\
\hline
\end{tabularx}
\end{table}

Table~\ref{tab:RepoLtCossim} shows the median of the cosine similarity of two issues that make up a link.
We see that \clo is the ``most similar'' link type, with an average~of 0.83. This makes sense as these links are created by a feature in JIRA that clones the text with only small changes made by the issue creators. 
We think that it might be interesting to investigate \clo links that have dissimilar issues, particularly if this was due to incremental changes over time.
\dup links are the second most similar, with around~0.35 cosine similarity score, on average, and a large gap to the similarity score of the \clo links. 
\rel links are the third most similar with an average score of~0.28.

Interestingly, \epi and \sub linked issues are very dissimilar. 
Considering that these are hierarchical relationships, one would assume that one of the issue's texts is somewhat contained in the corresponding other issues, which would result in a higher similarity.
This consistently low similarity together with the consistently high performance of these link types indicates that the model is not only learning textual similarities. 
Otherwise, the performance of \textit{non-links} would have been similarly high.
Additionally, the model is able to clearly differentiate \sub from \epi.

\begin{tcolorbox}[colback=gray!5!white,colframe=gray!75!black, size=small]
\underline{\textbf{Finding 4.}}
\clo and \dup pairs differ a lot in their textual similarity. \clo pairs are the most similar, followed by a wide gap and then by \dup.  Structural links \epi and \sub tend to connect token-dissimilar issues, as for \textit{non-links}. Since \epi and \sub have the highest prediction performance and can get differentiated from each other as well as \textit{non-links}, we can conclude that the model seems to be learning beyond the lexical similarity of issues.
\end{tcolorbox}
\vspace{6pt}

\begin{table}[t]
\setlength\tabcolsep{2.5pt}  
\centering
\captionsetup{skip=2pt}
    \caption{Median length of texts of linked issue on text-token-level across the JIRA repositories.}
    \label{tab:RepoLtTextLen}
    \begin{tabularx}{\columnwidth}{X  rrrrrrrrrr}
    \hline
\textbf{Repo}.  &  Rel. &  Dupl. &  Sub. &  Dep. &  Clo. &  Inc. &   Epic &  Blo. &  Cau. &  NL \\
\hline
\rowcolor{hellgrau}Apache        &   157 &      159 &     87 &   114 &   98 &        120 &   93 &  109 &  172 &       133 \\
Hyperledger   &   176 &      159 &     79 &      / &   92 &           / &   96 &  131 &     / &       105 \\
\rowcolor{hellgrau}IntelDAOS     &   268 &      398 &     90 &      / &  123 &           / &     / &  297 &     / &       142 \\
JFrog         &   152 &      172 &      9 &    70 &  260 &        195 &     / &     / &     / &       131 \\
\rowcolor{hellgrau}Jira          &   192 &      182 &     76 &   158 &  180 &        183 &     / &  161 &  244 &       168 \\
JiraEcosystem &   119 &      104 &     36 &   105 &  108 &         93 &   51 &   88 &  142 &        88 \\
\rowcolor{hellgrau}MariaDB       &   433 &      432 &     85 &      / &     / &        188 &   70 &  341 &  439 &       350 \\
Mojang        &   155 &      146 &       / &      / &  126 &           / &     / &  132 &     / &       112 \\
\rowcolor{hellgrau}MongoDB       &   165 &      154 &     68 &    96 &  100 &           / &   69 &     / &  174 &       119 \\
Qt            &   187 &      181 &     45 &   106 &  134 &           / &   89 &  216 &     / &       151 \\
\rowcolor{hellgrau}RedHat        &   135 &      136 &     74 &      / &  112 &         92 &     / &   97 &  153 &       110 \\
Sakai         &     174 &      153 &    106 &   150 &  123 &        127 &    / &  192 &    / &       143 \\
\rowcolor{hellgrau}SecondLife    &   163 &         / &     79 &   120 &  130 &        102 &     / &     / &     / &       160 \\
Sonatype      &   162 &      185 &     68 &   107 &     / &           / &  116 &     / &  184 &        50 \\
\rowcolor{hellgrau}Spring        &   173 &      179 &     58 &   107 &  124 &           / &   53 &     / &     / &       127 \\
\hline
\textbf{Mean} & 187 & 196 & \textbf{69} & 113 & 132 & 138 & \textbf{80} & 176 & \textbf{215} & 139  \\
\hline
\end{tabularx}
\end{table}

Table~\ref{tab:RepoLtTextLen} presents the median issue text length per repository, while 
Table~\ref{tab:RepoLtTextDiff} shows the median text length differences per repository. The Table shows the corresponding values for the studied link types.
We observe that \epi and \sub tend to connect  shorter issues than other link types, while issues connected via \cau, \rel, \dup, and \blo are longer.
Moreover, in line with the similarity results, we can observe that two issues linked by a \clo link are very similar in length, while other link types tend to have a higher difference in the word count.


\begin{table}[t]
\setlength\tabcolsep{2.5pt}  
\centering
\captionsetup{skip=2pt}
    \caption{Median difference of text length of two linked issues on text-token-level across the JIRA repositories.}
    \label{tab:RepoLtTextDiff}
    \begin{tabularx}{\columnwidth}{X  rrrrrrrrrr}
    \hline
\textbf{Repo}.  &  Rel. &  Dupl. &  Sub. &  Dep. &  Clo. &  Inc. &   Epic &  Blo. &  Cau. &  NL \\
\hline
\rowcolor{hellgrau}Apache        &    50 &       43 &     34 &    37 &    3 &         41 &  44 &   36 &   62 &        52 \\
Hyperledger   &    58 &       50 &     35 &      / &    2 &           / &  48 &   46 &     / &        46 \\
\rowcolor{hellgrau}IntelDAOS     &    91 &      146 &     30 &      / &    2 &           / &    / &  147 &     / &        59 \\
JFrog         &    44 &       49 &      5 &    18 &    2 &         61 &    / &     / &     / &        60 \\
\rowcolor{hellgrau}Jira          &     0 &       43 &     33 &    39 &    8 &         49 &    / &   43 &   68 &        54 \\
JiraEcosystem &    32 &       24 &     12 &    36 &    4 &         31 &  28 &   24 &   51 &        36 \\
\rowcolor{hellgrau}MariaDB       &   144 &      124 &     37 &      / &     / &         98 &  38 &  122 &  166 &       162 \\
Mojang        &    42 &       49 &       / &      / &   24 &           / &    / &   41 &     / &        35 \\
\rowcolor{hellgrau}MongoDB       &    58 &       47 &     22 &    36 &    1 &           / &  33 &     / &   63 &        49 \\
Qt            &    56 &       51 &     16 &    44 &    9 &           / &  40 &  143 &     / &        55 \\
\rowcolor{hellgrau}RedHat        &    42 &       37 &     27 &      / &    0 &         44 &    / &   35 &   52 &        46 \\
Sakai         &      52 &       38 &     36 &    48 &    3 &         41 &   / &   18 &    / &        48 \\
\rowcolor{hellgrau}SecondLife    &    51 &         / &     31 &    25 &   17 &         70 &    / &     / &     / &        54 \\
Sonatype      &    45 &       42 &     30 &    46 &     / &           / &  53 &     / &   64 &        23 \\
\rowcolor{hellgrau}Spring        &    52 &       50 &     23 &    38 &   44 &           / &  23 &     / &     / &        53 \\
\hline
\textbf{Mean} & 54 & 57 & 26 & 37 & \textbf{9} & 54 & 38 & 66 & \textbf{75} & 55 \\
\hline
\end{tabularx}
\end{table}

The link type \cau, performing worst, bears the highest difference in text lengths and the longest texts.
One possible explanation is that corresponding descriptions might contain a lot of information.
The BERT model cuts the text after a number of words and will likely not see the full text of all links. Particularly information of long stack traces might get lost.

\begin{table}[]
\centering
\captionsetup{skip=2pt}
\caption{Correlation coefficient and $p$-Value of macro F1-scores of the classifier to properties of the link types.}\label{tab:RepoLTPropCorr}
\begin{tabular}{lrr}
\hline
\textbf{Properties} &     \textbf{Correlation} &      \textbf{$p$-Value} \\
\hline
\rowcolor{hellgrau}\#Counts in Repos         &  0.2587 &  0.4704 \\
\#Difference          &  -0.4999 &  0.1412\\
\rowcolor{hellgrau}\#Length       &  \textbf{-0.6994} &  \textbf{0.0244} \\
\#Cosine Similarity      &  -0.1989 &  0.5817 \\
\hline
\end{tabular}
\end{table}

Table~\ref{tab:RepoLTPropCorr} shows the correlation of the analyzed link types properties, averaged across repositories, with the corresponding macro F1-score.
We observe that the only significant correlation is the text length.
We also see cosine similarity has a small correlation and is not significant, confirming that the model is not a simple similarity comparison model. 
Additionally, we observe that links that are mistaken by the model as \rel links often connect lengthier issues than issues of correctly classified links as well as issues of other mislabeled links (aside from \rel).

\begin{table*}[]
\centering
\captionsetup{skip=2pt}
\caption{Correlation coefficients of the F1-scores to properties of the link types. Significant correlations are in  bold.}\label{tab:LTPropCorr}
\begin{tabularx}{\textwidth}{X rrrrrrrrrr}
\hline
\textbf{Property} &     \multicolumn{10}{c }{Link Type}     \\
& Relate & Duplicate & Subtask & Depend & Clone & Incorporate & Epic & Block & Cause & Non-Link \\
\hline
\rowcolor{hellgrau}\#Support        &  0.3086   &  \textbf{0.5531} & 0.2399 & 0.5938 & 0.3900 & 0.5787 & -0.3888 & -0.0333 & 0.0821 & 0.2539 \\
Share in Training Data              &  \textbf{0.7561} &  \textbf{0.8765} & 0.4211 & \textbf{0.8938} & 0.1561 & \textbf{0.7795} & -0.0856 & 0.7133 & 0.1525 & 0.3486 \\
\rowcolor{hellgrau}Length           &  0.3514   &  -0.1507 & -0.4255 & -0.0631 & -0.2878 & -0.0004 & \textbf{-0.7852} & 0.6875 & 0.7447 & 0.1927 \\
Difference                          &  -0.0165  &  -0.1547 & -0.2916 & 0.2407 & \textbf{-0.8466} & -0.0744 & -0.7344 & 0.7381  & \textbf{0.7617} & 0.0793 \\
\rowcolor{hellgrau}Cosine Similarity& \textbf{0.5779}    &  0.3253  & -0.0222 & 0.1122 & \textbf{0.8451} & 0.0269 & -0.1673 & 0.2922 & -0.2032 & 0.3891 \\
\hline
\end{tabularx}
\end{table*}

Next, we explored individual link types in depth.
Table~\ref{tab:LTPropCorr} shows the individual correlations of each link type and its properties in the repositories.
For this, we excluded Mojang, as it is an outlier with three types and very good performance.

Unsurprisingly, we observe that the share of a link type in the training data correlates positively with the achieved performance, except for \epi. 
This observation is significant for \rel, \dup, \dep, and \inc. 
This is aligned with our previous observation that \dep and \inc require more data to be classified well.
The link type \epi depends on the length, the shorter a text, the better the classification.
The link type \clo correlates negatively with the text length difference and positively with the cosine similarity. This is the striking difference from other link types.
Last, \cau seems to depend on the difference in text lengths.

Finally, we calculated the Pearson correlation of the performance of link types to each other. 
That is, the existence of a link type might impact the performance of another link type, like \dep and \blo.
We found that \sub is harder to detect if the \epi is present in the repository with a correlation of -0.81.
Furthermore, the link type pairs \clo and \blo, \clo and \inc, and \blo and \inc strongly correlate positively ($\geq$ 0.9) with each other.

\begin{tcolorbox}[colback=gray!5!white,colframe=gray!75!black, size=small]
\underline{\textbf{Finding 5.}}
Text length of linked issues correlates negatively (-0.70) with the model performance. 
\epi and \sub links correspond to the shortest texts, while \cau, \rel, and \dup present longer texts.
Links mislabeled as \rel links tend to connect lengthier issues than a) correctly classified links and b) other mislabeled links.
Unsurprisingly, the share of the link type correlates positively for the link types \rel, \dup, \dep, and \inc.
\end{tcolorbox}

\section{Discussion}\label{sec:discussion}
We summarize our findings with possible interpretations  and implications for practice. We then discuss the importance of issue and link \textit{quality} and outline further research directions. Finally, we outline possible  threats to validity.

\subsection{Applicability of Typed Link Prediction in Practice}
While previous work has intensively studied duplication links (details in Section \ref{sec:rel-work}), this work is among the first to study the reliable prediction of issue links having different types (i.e. typed links).
The first resulting use case is to recommend to stakeholders missing issue links or to highlight incorrect link types. 
Our state-of-the-art BERT model achieved an average macro F1-score of~0.64 and a weighted F1-score of~0.73 across the~15 studied repositories.
These are promising results considering that the median amount of predicted classes is~7 (as opposed to, e.g., the simpler binary classification).

Our model is fairly general as it is end-to-end (without feature engineering) and it only uses the title and description of the issues as input features.
Additionally, even smaller repositories, like IntelDAOS (2599 links) showed a fair performance. Indeed, we found no significant correlation with the number of links (the training data).
The model works with mostly untouched data, only with as little basic preprocessing as possible. 
Stakeholders with more knowledge about the repository might fine-tune features based on internal workflows. 
Our model can further be optimized depending on the individual repository and project context.

The model was quite precise at predicting the hierarchical links \epi and \sub. It also showed similarly high accuracy for non-links. However, the model was not as good at identifying \dup links.
Counter-intuitively, it did not struggle to distinguish \clo from \dup links.
This is likely due to the way \clo links are created.
JIRA offers a feature to stakeholders to clone an issue. They create a new issue and start with all the properties of the cloned issue as default.
This explains the high cosine similarity of issues linked by clone.
In contrast, issues of a \dup link are usually created by two different contributors, who are often unaware of the other issue.
As \clo links are the only ones created automatically, a link prediction tool could also ignore them. 

Other link types, \dep, \inc, and \blo seem to need a critical mass to achieve good results.
The link type \cau has the lowest accuracy, which might be due to the fact that it is barely used or due to its length.
We truncated the texts, meaning that longer texts were cut. This affected link types which tend to be longer.
One reason for long texts is non-natural language fragments, such as stack traces or code.
It is also possible that stakeholders incorrectly link these issues with the \rel link type.
While manually reviewing the data, we observed that \rel links often contain stack traces or other pieces of non-natural language. 

There are three strategies that will likely improve the prediction performance and further increase the applicability in practice.
First, a tool could first detect the existence of a link and then its type with two different models.
The link detection model is trained on the dataset where all linked issues are in one class and randomly created \textit{non-links} in the other class.
The type detection model is trained on all links except the \rel links.
This alleviates the ensuing prediction problems of likely bad data quality for the \rel link type.

Second, a tool might predict categories of link types instead of a specific type.
With a median of 10.5 link types per repository, it is likely that some types are semantically related or partly redundant. 
Stakeholders in a project are better aware of which link types are similar. 
As our model is general to various types, no additional changes are needed, except for grouping the labels from certain types to a category.
Then, a stakeholder with knowledge about the project can choose the correct link type from the predicted category.

Third, to increase the model's applicability, the top~$k$ predictions can be presented to the stakeholder with multiple possible hits.
With $k=3$, we calculated that the average prediction performance of our results would increase by almost~18\%.


\subsection{Data Quality as Prerequisite for Correct Link Prediction}

Issue quality, in our case, text quality (missing information, ambiguity), directly impacts prediction quality as we only use the textual descriptions in linked issues.
For instance, the issues BE\mbox{-}213\footnote{\url{https://jira.hyperledger.org/browse/BE-213}} and BE-94\footnote{\url{https://jira.hyperledger.org/browse/BE-94}} in Hyperledger are linked as \clo and mislabeled as \epi by our model. 
Both of them only have a title (``Footer Components'' and ``Header Components'') but only an empty description.

Issue quality might be affected by the stakeholder creating the issue or the link (user, developer, analyst, product manager, etc.).
Zimmerman et al.~\cite{Zimmermann:2010:IEEESW} found that most issues reported by users miss important information for the developers, such as steps to reproduce or stack traces.
As users are often unaware of what makes good issue reports, they are likely to create lower-quality issues.
In contrast, analysts or developers are more aware that certain details are important to implement a requirement, fix a bug, or complete a certain task.




The link types \sub and \epi can be distinguished quite well with the text alone, likely due to the unique way they are treated in JIRA.
The main difference to other link types is that they have their own sections in the issue details. 
Thus, stakeholders might treat them more carefully.
Furthermore, \epi and \sub are often used to structure the issue tracking system. 
It is likely that \epi and \sub links are created with intent and care, usually by stakeholders deeply involved with the repository with a planning role. 
The link is likely created at issue creation time in an analysis and planning task. 
This all might improve the quality of the issue text as well as the correctness of the links.

In general, we think that the quality of the issues and the linking (i.e. the correctness of the link and its type) is only as good as the carefulness of the people who create them.
This is apparent in the seemingly rather average/low quality of \rel links, which is the most popular link type and most confused by the model at the same time.
We noticed, that other link types are often mislabeled as \rel links.
One explanation might be that the typed link prediction might suffer from low quality of \rel links.
Due to its general nature, \rel links are likely misused by stakeholders when they are uncertain whether other specific types are correct or more fitting.
Thus, \rel might contain a lot of data points that should have another label.
For instance, the issue ZOOKEEPER-3920\footnote{\url{https://issues.apache.org/jira/browse/ZOOKEEPER-3920}} in Apache has 
a \rel link to ZOOKEEPER-3466 and ZOOKEEPER-3828, while the comments discuss that this should be a \dup.

\subsection{Further Implications for Research}
We found repository and issue attributes that correlate with the prediction performance. Those were the coverage,  the issue-to-assignee ratio, as well as the text length of a link.
Furthermore, individual link type performance in repositories differs. \clo's performance depends on the textual similarity while other link types, such as \inc, \dep, \dup, and \rel depend on their share in the training data.
We think that two important factors, which require more extensive research are the link quality and the heterogeneity of a repository.

Coverage, the share of issues that are part of a link, correlates positively with the performance of the model and could be an indicator of higher data quality, making it easier for any model to learn.
A higher coverage implies that stakeholders try to link as many issues as possible or place a higher value on linking as a ``best practice'' to structure and manage the project knowledge.

The heterogeneity of an issue tracker is also an interesting factor in typed link prediction.
The issue-to-assignee ratio also correlates positively with the model performance, the more issues are assigned to one person, the better the model.
This makes sense: the fewer people involved, the less heterogeneity and more standardization can likely be observed in the repository.
We did not find any further significant correlation with the number of creators, reporters, or overall users.

Another indicator of heterogeneity is the number of projects in a repository, up to 646 in the case of Apache. 
We did not find a significant correlation between the number of cross-project links or the number of projects and the performance of the model on a repository. 
However, it might make sense to evaluate the prediction model per project or cluster of projects per repository instead, as links and their types inside a single project should be more homogeneous than in the whole repository.
Additionally, the usage of semantically overlapping link types (such as \dep and \blo) in a repository might point towards a certain heterogeneity. This could be an indicator for which issues or projects are ``similarly managed''. For instance,  one group might mainly use \blo for their issues while another group mainly \dep.


Finally, an interesting aspect of link and typed link prediction is identifying orphans, loners, or phantoms.
As coverage seems to be a factor for good typed link prediction, issues without any links to other issues or commits should be investigated.
Schermann et al.~\cite{Schermann:ICPC:2015} already examined and created a heuristic model to deal with links between issues and commits.

\subsection{Threats to Validity}
The data used for training and testing is largely representative of the way stakeholders use links and their types in practice.
The only restriction we placed was the 1\% share lower bound, as it is very conservative it should not introduce any bias that overestimates the quality of the model.
Practitioners with knowledge of the underlying processes and context can group their link types as they see fit, and the same model can still be applied.

As the labels, the link types, are made by humans, they can contain false positives.
We discuss likely quality issues and differences between the repositories that influence the quality of the model's prediction.
Additionally, we removed 1.1\% of the data as they contained multiple links with different types between two issues.
After reviewing them manually, we noticed that they were often conflicting. Moreover, as the percentage of multi-links was small, we chose to remove them since they do not warrant to use a multi-class multi-label classifier.
We also evaluated the approach on 15 different repositories and thus the approach is generalizable for repositories that use JIRA.
We did not evaluate the approach for Bugzilla and GitHub, which use fewer link types, thus the approach might produce different results for repositories that use other issue trackers.
As Bugzilla's default types (\rel, \dup, and \blo) are a subset of JIRA's default types, the model should achieve a similar performance in Bugzilla.
GitHub does not offer a specific link functionality. Thus, the model might be harder to apply there.

We added randomly created \textit{non-links} to the dataset. We chose to add as many as the mean number of other link types present to avoid a majority class which might bias the results.
With a higher amount of \textit{non-links} performance can change.

Finally, another possible threat is the evolution and changes over time in issue trackers.
If a repository has been in use for a long time, the implicit definition of link types can change too.
Issues themselves change over time and the links might not be updated accordingly and certain link types might fall out of favor over time.

\section{Related Work}\label{sec:rel-work}


With the rise of agile, requirements are often collected and tracked in issue trackers. 
This rise also led to a generation of large amounts of data in issue trackers, which are often too much to handle manually.
In a case study, Fucci et al.~\cite{Fucci:ESEM:2018} interviewed JIRA users and  found that information overload is one of their biggest challenges.
Interviewees of the study expressed the need for a requirements dependency identification functionality to reduce the overhead of discovering and documenting dependencies manually.
This paper provides a solid model that tackles this requirements dependency identification functionality.
Franch et al.~\cite{Franch:2021:RE} tackle the ensuing problems of agile practices in requirements elicitation and management with situational method engineering.
The collection and maintenance of requirements interdependencies (i.e. issue links) also face similar challenges.

Requirements Engineering research has largely studied another kind of dependency between issues/requirements and software artifacts: a topic known as traceability.
Deep learning has also been used for traceability.
Lin et al.~\cite{Lin:ICSE:2021} found that the traceability links between issue description to source code can be found with BERT which outperforms the traditional information retrieval methods, 
achieving an F1-score of 0.612 
and 0.729. 
Typed link prediction has similar application problems as traceability, such as poor quality in issue trackers, found by Merten et al.~\cite{Merten:REFSQ:2016}.
Additionally, Seiler et al.~\cite{Seiler:REFSQ:2017} conducted an interview study about the problems of feature requests in issue trackers and found that unclear feature descriptions and insufficient traceability are among the major problems in practice.
For typed link prediction we also found that heterogeneity of the repository is a problem.

Concerning issue link prediction, \dup is the most widely researched type, as duplication  detection is a tedious task~\cite{Cavalcanti:CSMR:2010} when curating issue trackers.
Deshmukh et al.~\cite{Deshmukh:ICSME:2017} proposed a single-channel siamese network approach with triplet loss which achieves an accuracy close to 90\% and a recall rate close to 80\%.
He et al.~\cite{He:ICPC:2020} proposed a dual-channel approach and achieved an accuracy of up to 97\%.
Rocha et al.~\cite{Rocha:ACCESS:2021} created a model using all ``Duplicate'' issues as different descriptions of the same issue and report a Recall@25 of~85\% for retrieval and an~84\% AUROC for classification.
All three works~\cite{Deshmukh:ICSME:2017, He:ICPC:2020, Rocha:ACCESS:2021} use the data set provided by Lazar et al.~\cite{Lazar:MSR:2014}, containing data mined from the four open-source Bugzilla systems: Eclipse, Mozilla, NetBeans, and OpenOffice.

There also exist studies researching other link types between issues and link usage.
Thompson et al.~\cite{Thompson:MSR:2016} studied three open-source systems and analyzed how software developers use work breakdown relationships between issues in JIRA. 
They observed little consistency in the types and naming of the supported relationships.
Li et al.~\cite{Li:APSEC:2018} examined the issue linking practices in GitHub and extracted emerging linking patterns. 
They cateogorized link types into 6 link type categories, namely: ``Dependent'', ``Duplicate'', ``Relevant'', ``Referenced'', ``Fixed'', ``Enhanced'', all the rarer link types were assigned the category ``Other''. 
They discovered patterns for automatic classification; ``Referenced'' links  usually refer to historic comments with important knowledge and that ``Duplicate'' links are usually marked within the day.
Tomova et al.~\cite{Tomova:ICSE:2018} studied seven open-source systems and reported that the rationale behind the choice of a specific link type is not always obvious.
The authors also found that \clo~links are indicative of textual similarity, issues linked through a \rel~link presented varying degrees of textual similarity and thus require further contextual information to be accurately identified.
From the varying degrees of textual similarity, we hypothesized that \rel links might be a jack-of-all-trades type.
Furthermore, while we found that some link types have distinct textual similarities, 
 they are not unequivocally identifiable only based on the textual similarity.

``Requires'' and ``Refines'' links were examined by Deshpande et al.~\cite{Deshpande:RE:2020}, who extracted dependencies 
on two industrial data sets achieving an F1-score of at least 75\% in both training sets.
\blo links were examined by Cheng et al.~\cite{Cheng:COMPSAC:2020} on the projects mined by Lazar et al.~\cite{Lazar:MSR:2014}.
They predicted the \blo link type with an F1-Score of 81\% and AUC of 97.5\%.


Most previous works view link types in isolation.
Recently, we evaluated state-of-the-art duplicate detection models on the same dataset used in this work and found that these models struggle with distinguishing \textit{Duplicate} from other link types~\cite{Lueders:MSR:2022}.
In this work, we focus on predicting multiple user-defined link types. We present and evaluate a BERT model for typed links prediction.


Nicholson et al.~\cite{Nicholson:AIRE_W:2020, Nicholson:AIRE:2021} also researched typed link prediction between issues of projects in Apache.
The first~\cite{Nicholson:AIRE_W:2020} analyzes the link types and tries to find patterns to predict missing links.
The second~\cite{Nicholson:AIRE:2021} evaluates several traditional machine learning approaches and achieves a weighted F1-score of about 0.563 up to 0.692 across the three projects HIVE, FLEX, and AMBARI.

Data quality, in this case, issue texts and link quality are essential for well-performing tools and smooth-running workflows.
We found that data quality affects typed link prediction as well.
In a similar line, Dalpiaz et al.~\cite{Dalpiaz:2021:RE} created an approach to improve the quality of user stories, a type of issue, by removing linguistic defects.
Link quality is directly affected by requirements quality. 
If issue descriptions are vague or contain other defects, links are harder to classify.

\section{Conclusion}\label{sec:conclusion}
Using BERT on the titles and descriptions of issue pairs, we achieved good performances to predict typed issue links on most studied repositories; and consistently excellent performance for predicting \epi and \sub links.
Our detailed analysis revealed that by better understanding the data and improving the issue quality (for the training and prediction) and the link quality (for the training) the prediction performance will likely get improved further -- particularly for correctly predicting the general purpose \rel links.
We  discussed strategies for increasing the performance and thus the model's applicability, particularly where  data quality is limited or heterogeneity in the issue tracker is high.
Future work should focus on the underlying data as well as the repository-, project-, and stakeholder-specific factors that might impact  the performance. 
For this user studies and qualitative research is needed to understand how and why stakeholders use the links and link types.

\section*{Acknowledgment}
We thank Lloyd Montgomery for collecting the dataset.
This work has been partly conducted within the Horizon 2020 project OpenReq, which is supported by the European Union under the Grant Nr. 732463.
\balance
\bibliographystyle{abbrv}
\bibliography{lib}

\end{document}